\documentclass[11pt]{article}

\usepackage{xr}
\makeatletter
\newcommand*{\addFileDependency}[1]{
  \typeout{(#1)}
  \@addtofilelist{#1}
  \IfFileExists{#1}{}{\typeout{No file #1.}}
}
\makeatother

\usepackage[english]{babel}
\usepackage[utf8x]{inputenc}
\usepackage[T1]{fontenc}
\usepackage{amsmath,amssymb}
\usepackage{algorithm}
\usepackage{mathtools}
\usepackage[noend]{algpseudocode}
\makeatletter
\def\BState{\State\hskip-\ALG@thistlm}
\makeatother

\usepackage[numbers,sort&compress]{natbib}
\usepackage[a4paper,top=3cm,bottom=2cm,left=3cm,right=3cm,marginparwidth=1.75cm]{geometry}
\usepackage{graphicx}
\usepackage{subcaption}
\usepackage{float}
\usepackage{accents}
\usepackage{fancybox}
\usepackage{geometry}
\allowdisplaybreaks
\usepackage{authblk}

\title{\LARGE{\textbf{A tuned mass amplifier for enhanced haptic feedback}}}
\author[1,2]{Sai Sharan Injeti\thanks{ssharaninjeti@gmail.com}}
\author[1]{Ali Israr}
\author[1]{Tianshu Liu}
\author[1]{Yi\u{g}it Meng\"{u}\c{c}}
\author[1]{Daniele Piazza}
\author[1]{Dongsuk D. Shin\thanks{shindaniel@fb.com}}
\affil[1]{\small{Facebook Reality Labs Research, Redmond, WA, USA}}
\affil[2]{\small{Division of Engineering and Applied Science, California Institute of Technology, Pasadena, CA, USA}}
\date{}
\usepackage{color}
\begin{document}
\maketitle
\begin{abstract}
\normalsize
\bigskip
Vibro-tactile feedback is, by far the most common haptic interface in wearable or touchable devices. This feedback can be amplified by controlling the wave propagation characteristics in devices, by utilizing phenomena such as structural resonance. However, much of the work in vibro-tactile haptics has focused on amplifying local displacements in a structure by increasing local compliance. In this paper, we show that engineering the resonance mode shape of a structure with embedded localized mass amplifies the displacements without compromising on the stiffness or resonance frequency. The resulting structure, i.e., a \textit{tuned mass amplifier}, produces higher tactile forces (7.7 times) compared to its counterpart without a mass, while maintaining a low frequency. We optimize the proposed design using a combination of a neural network and sensitivity analysis, and validate the results with experiments on 3-D printed structures. We also study the performance of the device on contact with a soft material, to evaluate the interaction with skin. Potential avenues for future work are also presented, including small form factor wearable haptic devices and remote haptics.  
\end{abstract}

\section{Introduction}
Interaction between humans and computational devices mainly relies on visual or auditory feedback. With most interactive devices, the user's vision are overloaded, which makes utilizing haptic feedback an important sensory interface. For example, haptic feedback can be used for navigation when the user is using visual feedback to avoid obstacles \cite{ertan1998,lindeman2006}. It is also a useful interface in applications such as robot assisted minimally invasive surgery \cite{okamura2009} and prostheses for rehabilitation \cite{plauche2016}. Haptic feedback further enables sensing fine textures and shape changes that may not be visually perceivable \cite{johansson2009}. The first type of haptic interface includes touch-based feedback, that relies on the high density of mechanoreceptors on fingertips \cite{johansson2009,chortos2016}. It is more challenging to design for the second kind of haptic interface, wearables, as the skin around such a device tends to have fewer receptors \cite{withana2018}. 

The three main categories of wearable haptic displays are force feedback, electro-tactile feedback and vibro-tactile feedback devices \cite{kurita2014}. A force feedback device typically uses an exoskeleton, which is capable of exerting large forces, but tends to be bulky and often requires user-specific calibration \cite{yang2004}. Electro-tactile feedback devices rely on surface electrodes that directly stimulate the muscles themselves to contract, which sometimes causes pain to users and hence requires a challenging design and calibration to avoid lesions \cite{yu2019}. Vibro-tactile feedback has shown to be an effective way to produce tactile sensations by using low frequency mechanical forces. Carefully arranging vibration actuators within a device allows for a range of tactile spatial resolutions \cite{reardon2020,hudin2015,bilal2020,paneels2013}. A fine spatial resolution requires localized vibration on the structure/ device to stimulate mechanoreceptors at specific regions. 

The first way to create local vibrations is by utilizing time-reversal for elastodynamic waves \cite{reardon2020,hudin2015}. Such techniques often result in the need for multiple actuators, which can result in a bulky device. A second way to induce local vibrations is by utilizing local resonance, where a single actuator can be vibrated at different frequencies to locally excite compliant regions in the device \cite{bilal2020,paneels2013}. So far, this mechanism has proven to be effective in amplifying the vibration displacements from thin and weak actuators at locally distinct regions in a structure, leading to the possibility of low form factor wearable devices. However, achieving large amplifications in displacements together with high haptic forces at low frequency has remained challenging. Typical methods to amplify local displacements at low frequency rely on softening the structure locally, and as a result compromise on local stiffness. This is the focus of our work. In this paper we present a way to amplify local displacements in a structure at low resonance frequencies for haptics, while still maintaining high structural stiffness, hence demonstrating our \textit{tuned mass amplifier} (TMA). 

\begin{figure}
    \centering
    \includegraphics[width=1.\textwidth]{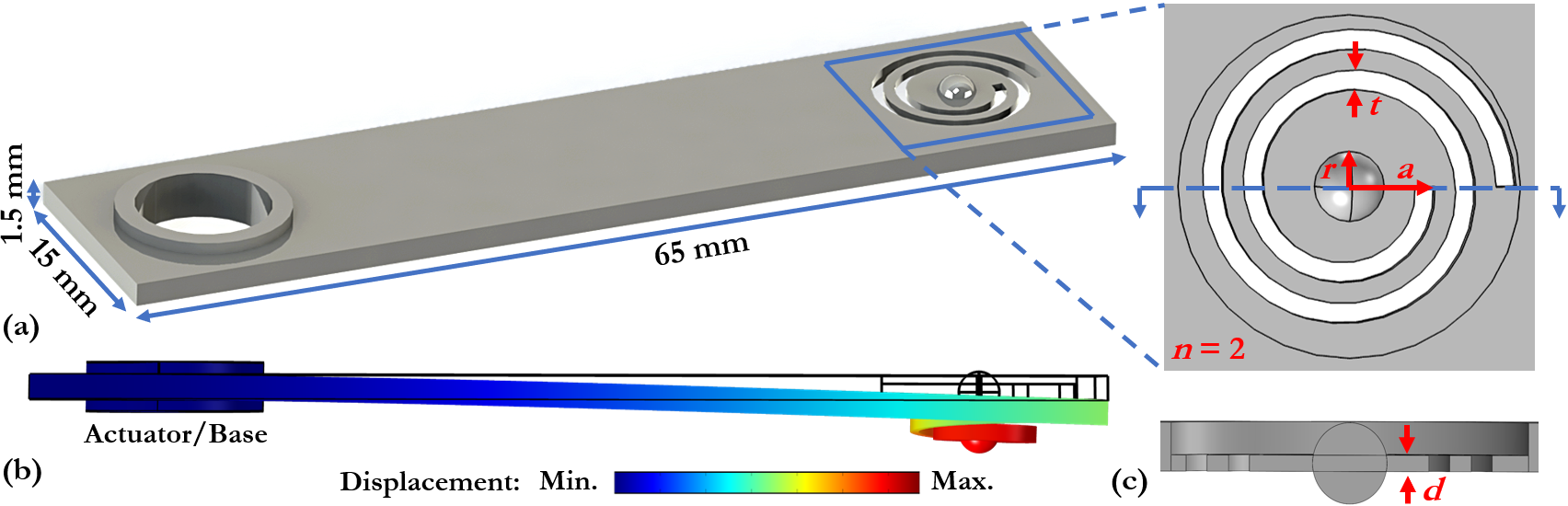}
    \caption{\textit{\textbf{(a)} A tuned mass amplifier consisting of a stiff primary structure (a cantilever), a secondary compliant structure (a circular region with spiral cut) with an embedded heavy spherical mass. \textbf{(b)} A mode of flexural resonance showing the compliant region resonating in-phase with the rest of the cantilever. \textbf{(c)} Close up of the compliant region containing the variable design parameters, indicated in red.}}
    \label{fig:1}
\end{figure}

We demonstrate the concept where a structure can be resonated such that a primary and secondary structure move in-phase to amplify displacements and stiffness, and a local mass can be used to ensure the phenomenon happens at low frequency for haptics. As an example of this concept, our proposed device shown in Fig. \ref{fig:1}(a) consists of a primary structure (a cantilever), with a secondary compliant structure made from an Archimedian spiral cut \cite{bilal2020}, with an embedded mass that is several times more dense compared to the material used to make the remaining structure. We tune the parameters describing this geometry such that the resonance mode shape consists of the cantilever and compliant spiral displacing in-phase (Fig. \ref{fig:1}(b)), hence amplifying both local out-of-plane displacement and stiffness. The embedded mass ensures that the resonance frequency remains low. This combination of high static stiffness and large resonance displacements leads to high tactile forces. We design the structure to achieve a desired resonance mode shape, frequency and static stiffness, by utilizing a neural network to learn the required outputs from finite element analysis and then solve an optimal design problem using a sensitivity analysis. We perform dynamic and static experiments on 3-D printed structures and validate our analysis with measurements of the resonance frequency, displacements, and local static stiffness. We quantify the force feedback and the effect of skin with experiments where the device is in contact with a soft material. Note that the design of the base where the actuator is mounted (Fig. \ref{fig:1}(b)) is guided by the actuator we use in experiments, as detailed later in the paper.

\section{Optimal design}

We model the dynamic and static response of the structure using finite element analysis in COMSOL. For a non-dissipative linear elastic medium, let $\underaccent {\tilde}K\left(\underaccent {\tilde}\chi\right)$ and $\underaccent {\tilde}M\left(\underaccent {\tilde}\chi\right)$ represent the assembled stiffness and mass matrices, where $\underaccent {\tilde}\chi$ is the vector of parameters that describes the topology of the structure. For a harmonic excitation at the base as indicated in Fig. \ref{fig:1}(b), the eigenvalue problem
\begin{equation}\label{eq:dynamic}
    \left( \underaccent {\tilde}K\left(\underaccent {\tilde}\chi\right) - \left(\omega(\underaccent {\tilde}\chi)\right)^2 \underaccent {\tilde}M\left(\underaccent {\tilde}\chi\right) \right) \underaccent {\tilde}U\left(\underaccent {\tilde}\chi\right) = 0,
\end{equation}
indicates modes of resonance of the structure \cite{hughes2012}. Each mode shape $\underaccent {\tilde}U\left(\underaccent {\tilde}\chi\right)$ represents the normalized displacements at the nodes of the finite elements in the structure at a resonance frequency $\underaccent {\tilde}\omega\left(\underaccent {\tilde}\chi\right)$. For a static problem where the base of the structure is fixed and a point load of magnitude $f$ is applied at the center of the spiral 
\begin{equation} \label{eq:static}
    \underaccent {\tilde}K\left(\underaccent {\tilde}\chi\right) \underaccent {\tilde}V\left(\underaccent {\tilde}\chi\right) = \underaccent {\tilde}F,
\end{equation}
where $\underaccent {\tilde}F$ (the force vector) contains only one non-zero element of magnitude $f$ corresponding to the node at the center of the spiral. $\underaccent {\tilde}V\left(\underaccent {\tilde}\chi\right)$ represents the small deformation static displacement in the structure for the applied load. In vectors $\underaccent {\tilde}U$ and $\underaccent {\tilde}V$, let the $i^{th}$ and $j^{th}$ elements represent the displacement of the node at the center of the spiral and at the upper-right tip of the cantilever, respectively (Fig. \ref{fig:1}(a)).

\subsection{Learning the model}

We define three key mechanical properties that can influence haptic feedback. The first property is the frequency of excitation $\left(\omega(\underaccent {\tilde}\chi)\right)$ at the base that produces a desired (coupled) mode of resonance as illustrated in Fig. \ref{fig:1}(b). 
Second, we introduce a property for a measure of mode localization $\left(L(\underaccent {\tilde}\chi)\right)$, which we define as the ratio of displacement at the center of the spiral to the tip of the cantilever for the coupled mode shape (i.e. $L=U_i/U_j$). 
Third, we aim to design for the local stiffness of the structure at small deformations $\left(S(\underaccent {\tilde}\chi)\right)$ subject to a unit point load at the center of the spiral ($S=1/V_j$). A locally stiffer structure would result in higher tactile forces for a given local displacement. We pick the base material to be Nylon 12, which we use to 3-D print these structures. 
The mass at the center of the spiral is made of tungsten carbide, which is more than 16 times as dense as Nylon 12. The linear elastic material properties for the materials used are listed in table \ref{tab:1} \cite{knoop2015, lin2009}. 
We measure the density from the ratio of mass to volume of a 1 cm side cube. We determine its Poisson's ratio from analytically verifying the first resonance frequency of a solid cantilever beam made of Nylon 12, with length 65 mm, breadth 15 mm and height 1.5 mm, with finite elements. We obtain the Young's modulus of elasticity and Poisson's ratio from the bulk and shear moduli of Tungsten Carbide reported in \cite{lin2009}, and its density from the ratio of mass to volume of a spherical ball with a diameter of 3 mm.

\begin{table}
\begin{center}
\caption{Linear elastic material properties} 
\label{tab:1}
{\renewcommand{\arraystretch}{1.5}
\begin{tabular}{ |c|c|c|c| } 
\hline
\textbf{Material} & \textbf{Young's modulus} ($E$) & \textbf{Poisson's ratio} ($\nu$) & \textbf{Density} ($\rho$) \\
\hline
Nylon 12 & 1.2 GPa & 0.34 & 983 kg/m$^3$ \\ 
Tungsten Carbide& 702 GPa & 0.20 & 16,000 kg/m$^3$\\ 
\hline 
\end{tabular}
}
\end{center}
\end{table}

The optimal design of the structure to yield a desired set of mechanical properties ($\omega$, $S$ and $L$) can be time consuming if we purely rely on the finite element simulation to estimate them at each step of any optimization algorithm. The time taken increases drastically, when there is a need to design for multiple devices each with a distinct set of mechanical properties. For example, if the number of design iterations in one optimization algorithm to design for a set of mechanical properties is $\sim$10,000 and we are interested in designing 200 different devices each with unique properties, we are looking at a total of $\sim$2 million finite element simulations, which is extremely expensive.

\begin{figure}
    \centering
    \includegraphics[width=1.\textwidth]{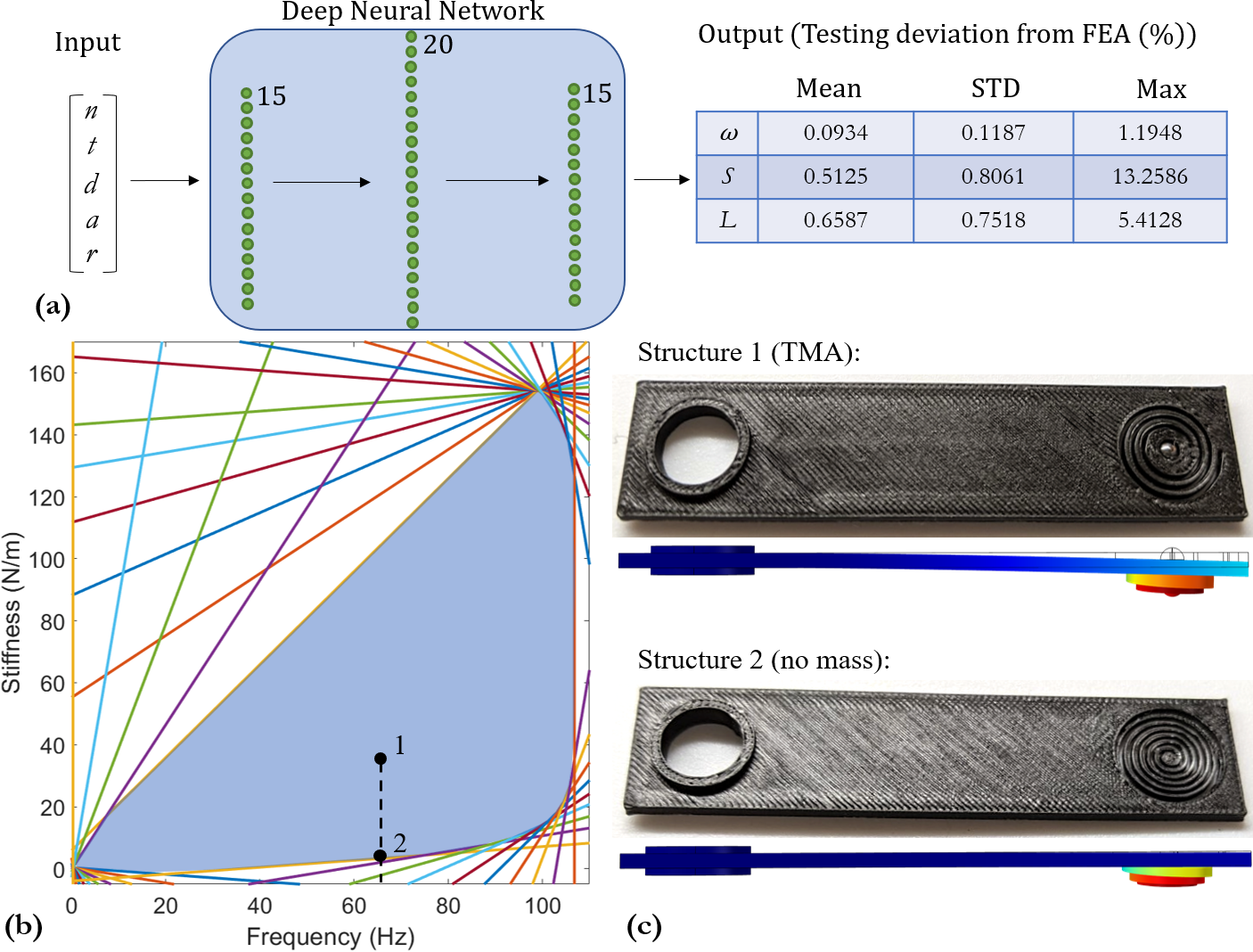}
    \caption{\textit{\textbf{(a)} Architecture of the deep neural network to learn the mechanical properties given the design parameters. The table indicates the performance of the network. \textbf{(b)} Bounds on frequency and stiffness for design parameters within the range in the table \ref{tab:2}. \textbf{(c)} Structure 1 (TMA with coupled resonance) and structure 2 (without an embedded mass and localized resonance at the same frequency) are tested in experiments, and their properties are indicated in (b).}}
    \label{fig:2}
\end{figure}

We train a neural network to learn the finite element simulation for a given set of design parameters, to improve the computational time to estimate the properties. We pick five design parameters to represent $\underaccent {\tilde}\chi=[n,\ t,\ d,\ a,\ r]^T$, that do not change the form factor of the device in Fig. \ref{fig:1}. These are the number of turns in the spiral cut ($n$), thickness of the spiral cut ($t$), depth of the spiral ($d$), smallest outer radius of the spiral ($a$) and the radius of the center mass ($r$), respectively. Note that we fix the length, breadth and height of the cantilever to fix the form factor of the device. The fixed dimensions and variable design parameters are indicated in blue and red in Fig. \ref{fig:1}(a) and (c), respectively. We pick bounds for each variable design parameter as indicated in table \ref{tab:2} and generate 21,000 geometries that are uniformly distributed within these bounds. We ensure that each generated point represents a structure where the minimum feature size is larger than $\delta=0.1$ mm, to aid meshing. We train a deep neural network with three hidden layers (of 15 nodes, 20 nodes and 15 nodes, respectively) that takes the $5 \times 1$ design vector as the input and learns the mechanical properties calculated using finite elements. We train 3 such separate networks to estimate each property, $\omega$, $S$ and $L$. We use 17,000 data points for training, and 2,000 data points each for testing and validation. Each data point represents an input vector containing the design variables and output containing the frequency of resonance, stiffness or localization. We use MATLAB's machine learning toolbox, where we minimize the mean squared error, using a gradient descent algorithm. The activation function used is the rectified linear unit (ReLU). A schematic of the network along with its performance against FEA is indicated in Fig. \ref{fig:2}(a).

\subsection{Calculation of bounds}

We formulate the following optimal design problem to calculate the bounds on frequency and stiffness of the device,
\begin{equation} \label{eq:opt_bounds}
\begin{aligned}
\min_{\underaccent {\tilde}\chi} \quad & \gamma_1 \omega(\underaccent {\tilde}\chi) + \gamma_2 S(\underaccent {\tilde}\chi)\\
\textrm{s.t.} \quad & \underaccent {\tilde}A\underaccent {\tilde}\chi \leq \underaccent {\tilde}b,\\
  &f_1(\underaccent {\tilde}\chi) \geq \delta,\\
  &f_2(\underaccent {\tilde}\chi) \geq \delta.\\
\end{aligned}
\end{equation}
Solving problem \eqref{eq:opt_bounds} for several pairs of real numbers $\gamma_1$ and $\gamma_2$ such that $|\gamma_1|+|\gamma_2|=1$, and taking the intersection of all half planes $\gamma_1 \omega(\underaccent {\tilde}\chi) + \gamma_2 S(\underaccent {\tilde}\chi) \geq \mathcal{O}^{min}$ gives us the outer envelope of possible values of $\omega$ and $S$. It is an outer bounds since all attainable values lie within this intersection. Similarly, replacing $S$ with $L$ in problem \eqref{eq:opt_bounds} gives us bounds on frequency versus localization. The linear inequality constraint $\underaccent {\tilde}A\underaccent {\tilde}\chi \leq \underaccent {\tilde}b$ is formulated to satisfy the bounds of parameters in table \ref{tab:2}. The constraint $f_1(\underaccent {\tilde}\chi) \geq \delta$ limits the minimum solid thickness of the spiral to 0.1 mm. This function can be approximated as $f_1(\underaccent {\tilde}\chi)= \dfrac{(a' - a) - ([n]+1)t}{[n]}$, where $a'=5.85$ mm is the fixed largest outer radius of the spiral and $[n]$ is the greatest integer less than or equal to $n$. Note that $f_1$ is a conservative estimate of the minimum feature size. The constraint $f_2(\underaccent {\tilde}\chi) \geq \delta$ limits the maximum radius of the tungsten carbide ball. We have that $f_2(\underaccent {\tilde}\chi)= a-t-r$. We solve problem \eqref{eq:opt_bounds} using gradient descent optimization. The sensitivities with respect to each design parameter are numerically calculated by perturbing the design variable along each orthogonal direction and measuring the change in the objective. To avoid local minima, we perform each optimization starting from several feasible initial guesses and finalize the best value. The results are indicated in Fig. \ref{fig:2}(b). The blue shaded area enclosed by the half planes represents the region within which all attainable values of resonance frequency $\omega$ (of the first mode) and local stiffness $S$ lie. A similar approach to calculating the bounds between such highly non-linear properties can be found in \cite{injeti2019}.

\begin{table}
\begin{center}
\caption{Bounds on design variables} 
\label{tab:2}
{\renewcommand{\arraystretch}{1.5}
\begin{tabular}{ |c|c|c| } 
\hline
\textbf{Bounds} & \textbf{Minimum} & \textbf{Maximum}\\
\hline
$n$ & 2 & 6 \\ 
$t$ & 0.2 mm & 0.5 mm \\ 
$d$ & 1 mm & 1.5 mm \\
$a$ & 1 mm & 3 mm \\
$r$ & 0 mm & 3 mm \\
\hline 
\end{tabular} 
}
\end{center}
\end{table}

\subsection{Design of samples}

To illustrate the concept of a TMA, we pick a low frequency of $\omega$=65 Hz and engineer two structures, one with a coupled mode of resonance and a local mass, and the other without an embedded mass and localized mode of resonance at the spiral. The former yields our tuned-mass amplifier (with a design vector ${\underaccent {\tilde}\chi}_1$) and possesses high stiffness. The latter yields a representative structure for the state-of-the-art method that solely maximizes localization and compliance to amplify haptic feedback \cite{bilal2020} (with a design vector ${\underaccent {\tilde}\chi}_2$). We pick a frequency of 65 Hz as the human perception drops over 100 Hz \cite{israr2007}. We formulate the following optimal design problem to solve for the topologies of the two structures
\begin{equation} \label{eq:opt_samples}
\begin{aligned}
\min_{{\underaccent {\tilde}\chi}_1,{\underaccent {\tilde}\chi}_2} \quad & \left( \omega({\underaccent {\tilde}\chi}_1)- 65\ \textrm{Hz}\right)^2 + \left( \omega({\underaccent {\tilde}\chi}_2)- 65\ \textrm{Hz}\right)^2 + \textrm{sin}^4\left(\dfrac{\pi {\underaccent {\tilde}\chi}_1 . \underaccent {\tilde}v}{0.25\ \textrm{mm}}\right) + \textrm{sin}^4\left(\dfrac{\pi {\underaccent {\tilde}\chi}_2 . \underaccent {\tilde}v}{0.25\ \textrm{mm}}\right) \\ 
& + \Phi_p \left( 2.5- L({\underaccent {\tilde}\chi}_1)\right) + \Phi_p \left( L({\underaccent {\tilde}\chi}_1) - 3 \right) + \Phi_p \left( 8- \frac{S({\underaccent {\tilde}\chi}_1)}{S({\underaccent {\tilde}\chi}_2)}\right)\\
\textrm{s.t.} \quad & \underaccent {\tilde}A{\underaccent {\tilde}\chi}_1 \leq \underaccent {\tilde}b,\ \underaccent {\tilde}A{\underaccent {\tilde}\chi}_1 \leq \underaccent {\tilde}b,\\
  &f_1({\underaccent {\tilde}\chi}_1) \geq \delta,\ f_1({\underaccent {\tilde}\chi}_2) \geq \delta,   \\
    &f_2({\underaccent {\tilde}\chi}_1) \geq \delta,\ f_2({\underaccent {\tilde}\chi}_2) \geq \delta,   \\
&\underaccent {\tilde}v=(0\ 0\ 0\ 0\ 1)^T.   \\
\end{aligned}
\end{equation}
The first two terms in the objective of problem \eqref{eq:opt_samples} ensure that the resonance frequencies of the two optimal designs are close to the desired value. The third and fourth terms ensure that the radius ($r$) of the Tungsten Carbide ball is a multiple of 0.25 mm, due to the availability of precision ball bearings in such denominations. The power of the trigonometric functions are chosen to avoid terms with much smaller sensitivities than others at a local minimum. We define a penalty function $\Phi_p(x)=\dfrac{1}{p}(e^{px}-1)$, with $p$ being large (we take $p=50$ for our calculations) \cite{jayswal2014}. Notice that $\Phi_p(x)$ is non-positive and low in magnitude when $x\leq 0$, and is positive and large otherwise. The fifth and sixth terms in the objective ensure that the optimal design for the first structure (indexed by 1) produces a localization between 2.5 and 3, ensuring a coupled mode of resonance. The last term in the objective drives the stiffness of the first structure to be much larger than the second, ultimately forcing a localized mode of resonance in the second structure (indexed by 2). This is because lower stiffness occurs when the spiral region is much softer than the cantilever, which also results in a localized mode of resonance. When a point load at the center of the spiral displaces both the spiral and cantilever, this results in higher stiffness as well as a coupled mode of resonance. The optimal values, ${\underaccent {\tilde}\chi}_1^*=[2.99, 0.45\textrm{ mm}, 1.50\textrm{ mm}, 2.70\textrm{ mm}, 1.25\textrm{ mm}]^T$ and ${\underaccent {\tilde}\chi}_2^*=[6.00, 0.45\textrm{ mm}, 1.00\textrm{ mm}, 1.02\textrm{ mm}, 0.00\textrm{ mm}]^T$. Table \ref{tab:3} compares the result predicted for these optimal designs using the neural network, with FEA. The values of the properties are in agreement, and we clearly notice structure 1 displays a much higher stiffness compared to structure 2 at the same resonance frequency. We 3-D print the structures using fused deposition modeling (FDM) with Nylon 12 (Fig. \ref{fig:2}(c)). The tungsten carbide ball is press-fit into a cavity in structure 1. The properties of these structures and their mode shapes of resonance are indicated in Fig.\ref{fig:2}(b) and (c), respectively. 

\begin{table}
\begin{center}
\caption{Properties of the optimal designs from analysis} 
\label{tab:3}
{\renewcommand{\arraystretch}{1.5}
\begin{tabular}{ |c|c|c| } 
\hline
\textbf{Property} & \textbf{Neural Network} & \textbf{FEA} \\
\hline
$\omega({\underaccent {\tilde}\chi}_1^*)$ & 65.8789 Hz & 64.8067 mm  \\
$L({\underaccent {\tilde}\chi}_1^*)$ & 2.6727 & 2.7197  \\
$S({\underaccent {\tilde}\chi}_1^*)$ & 34.9035 N/m & 34.8287 N/m  \\ 
\hline 
$\omega({\underaccent {\tilde}\chi}_2^*)$ & 65.3809 Hz & 64.3779 mm  \\ 
$L({\underaccent {\tilde}\chi}_2^*)$ & 19.0052 & 19.0030  \\
$S({\underaccent {\tilde}\chi}_2^*)$ & 3.9912 N/m & 3.9350 N/m  \\ 
\hline
\end{tabular} 
}
\end{center}
\end{table}

\section{Experiments}

\subsection{Validation of static and dynamics properties}

\begin{figure}
    \centering
    \includegraphics[width=1.\textwidth]{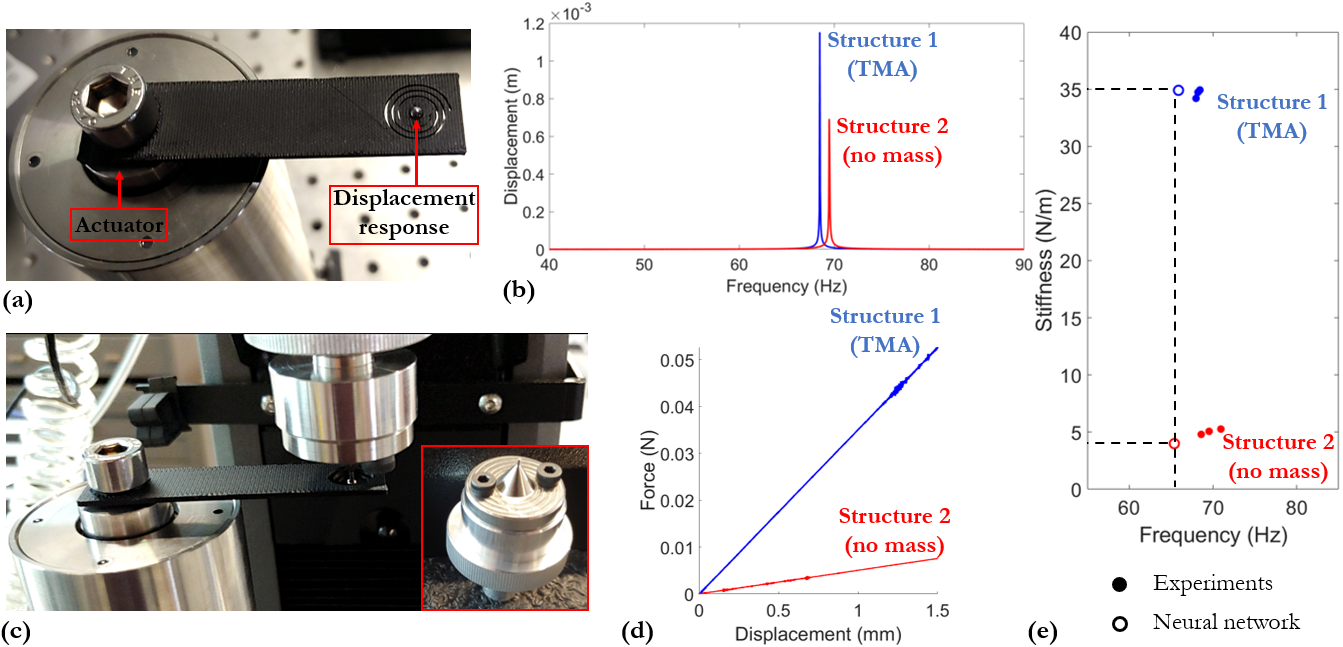}
    \caption{\textit{\textbf{(a)} Setup showing the TMA attached to a piezoelectric stack actuator for vibration measurements. \textbf{(b)} Displacement response in the two structures excited at their respective resonance frequencies. \textbf{(c)} Setup showing the TMA loaded on an Instron machine. The inset shows the sharp indenter used. \textbf{(d)} Force vs displacement response of the two structures for small deformation. \textbf{(e)} Comparison between the properties measured in experiments with the analysis.}}
    \label{fig:3}
\end{figure}

In order to validate the mechanical properties estimated from the analysis, we first measure the desired resonance frequencies of the two optimal designs. We use a piezoelectric stack actuator (model P-235) as the vibration source, that we attach to our sample at its base using an M-8 screw (Fig. \ref{fig:3}(a)). Due to the extremely high stiffness of the actuator, the boundary conditions at the base of our structures resemble our analysis, i.e. displacement boundary conditions. We measure the displacement response at the center of the spiral in each case using a laser vibrometer. We perform the experiments on three samples for each structure, to account for variability in the 3-D printing process. We provide sinusoidal input vibration with a frequency sweep from 40 Hz to 90 Hz, to identify the first resonance frequency in each structure. To identify the resonance frequency, we take the Fourier transform of the displacement versus time data as measured by the vibrometer, and locate the frequency corresponding to the peak in displacement. The results are indicated in table \ref{tab:4}. Notice that the mean values agree very well with our analysis from table \ref{tab:3}. 

\begin{table}
\begin{center}
\caption{Properties of the optimal designs from experiments} 
\label{tab:4}
{\renewcommand{\arraystretch}{1.5}
\begin{tabular}{ |c|c|c| } 
\hline
\textbf{Property} & \textbf{Structure 1 (TMA)} & \textbf{Structure 2 (no mass)} \\
\hline
Resonance frequency & 68.203 $\pm$ 0.235 Hz & 69.686 $\pm$ 1.182 Hz  \\
Stiffness & 34.622 $\pm$ 0.378 N/m & 5.044 $\pm$ 0.233 N/m  \\ 
Displacement amplification & 135.378 $\pm$ 28.116 & 84.307 $\pm$ 13.049 \\
\hline
\end{tabular} 
}
\end{center}
\end{table}

At their respective measured resonance frequencies, we excite each structure for 15 seconds and measure the displacement versus time response at the center of the spiral. The Fourier transform of the data is indicated in Fig. \ref{fig:3}(b) for representative samples in each topology. The amplitude of vibration at the base is (9.263 $\pm$ 1.031) $\times 10^{-6}$ m. Notice that the displacement amplification at the spiral is much larger for the tuned mass amplifier (structure 1), when compared to structure 2 that does not feature a local mass. This can be attributed to the coupled resonance that is engineered in structure 1. Structure 2 amplifies displacements solely due to the high compliance of the spiral. The displacement amplifications at resonance in each structure is summarized in table \ref{tab:4}.

In order to measure the local stiffness of a structure at the center of the spiral, we use the setup shown in Fig. \ref{fig:3}(c). The piezoelectric stack actuator is fixed to a table (stationary) and the sample is screwed into the actuator at its base. We load the center of the spiral on an Instron machine, with a sharp indenter that has a tip diameter of 1 mm. We displace the indenter by 1.5 mm at a low strain rate of 0.01 mm/s to measure the quasi-static small deformation response. The measured force vs displacement curve for representative samples in each case is shown in Fig. \ref{fig:3}(d). Notice the linear response of the structure for small deformation, indicating that a linear elastic material model is sufficient to model small deformation vibrations. The local stiffness is measured as the slope of the curve near the origin, and the results are tabulated in table \ref{tab:4}. As expected, the tuned mass amplifier displays a much higher stiffness than the other structure. 

Fig. \ref{fig:3}(e) compares the experimentally measured stiffness and resonance frequency of the three samples with either topology, to the values estimated from the neural network. We see great agreement between our analysis and experiments and 
clearly demonstrate the engineered phenomenon: a large local stiffness and displacement amplification without any change in resonance frequency. 

\subsection{Performance in contact}

\begin{figure}
    \centering
    \includegraphics[width=1.\textwidth]{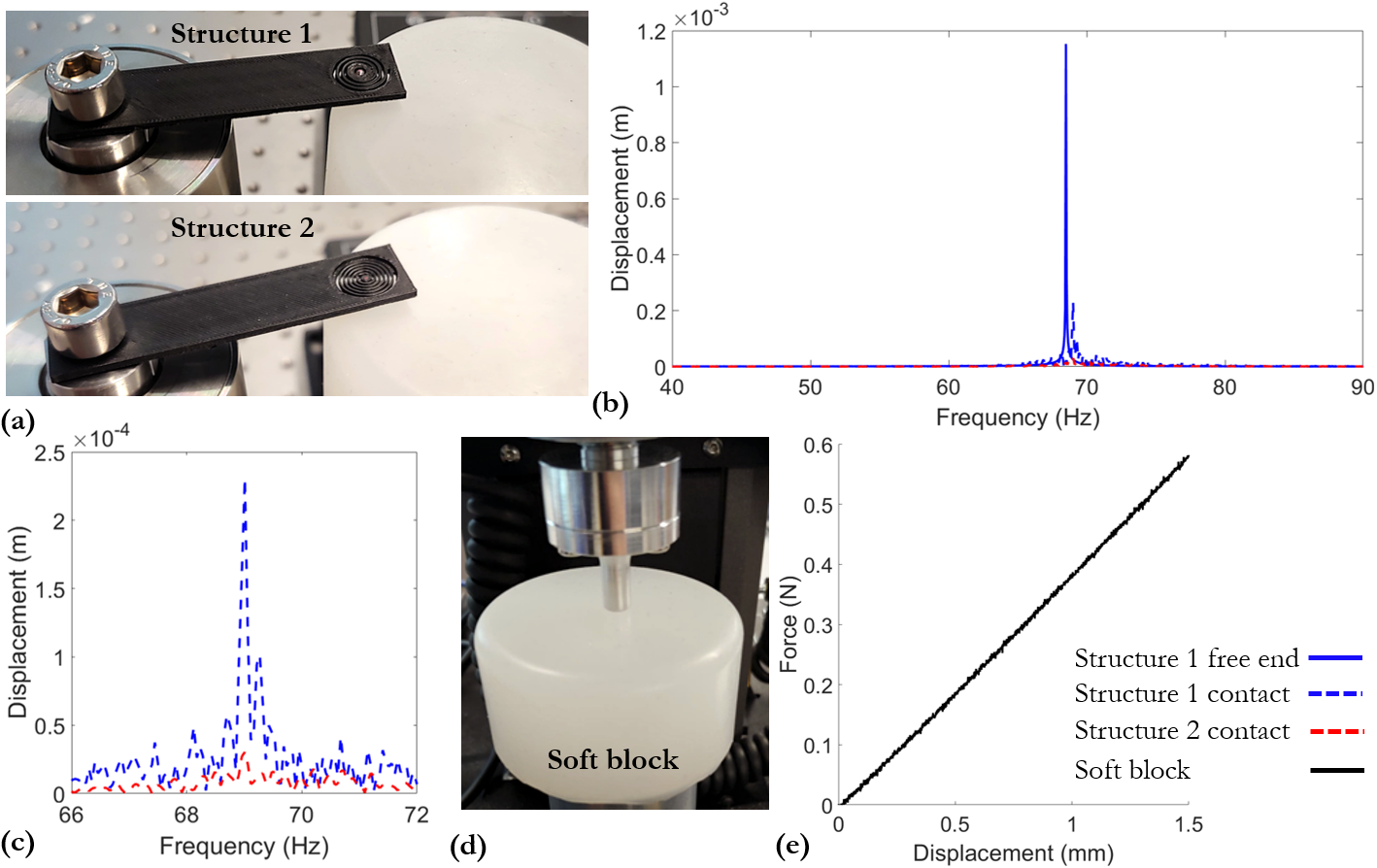}
    \caption{\textit{\textbf{(a)} Vibration test setup showing structures 1 and 2 in contact with the soft block. \textbf{(b)} Displacement response in the structures (structure 1 without and with contact, and structure 2 with contact) excited at their respective resonance frequencies. \textbf{(c)} Close up of (b) showing the comparison between structures 1 and 2 with contact. \textbf{(d)} Setup showing the soft block loaded on an Instron machine. \textbf{(e)} Force vs displacement response of the soft material for small deformation.}}
    \label{fig:4}
\end{figure}

The device would ultimately be used in contact with skin, and it is important to understand the vibration response with a contact boundary condition to evaluate its haptic performance. The increase in stiffness and displacement would translate into a higher tactile force when the device comes in contact with skin, as felt by the authors. To validate and quantify this, we measure the displacements and tactile stresses by repeating the vibration experiments, with the device now in contact with a soft material block. We use a platinum-catalyzed silicone block (Ecoflex$^{\textrm{TM}}$ 00-30) in contact with the spiral on each structure, to replicate the effect of touch (Fig. \ref{fig:4}(a)). The cured rubber is soft, strong and elastic for large deformation, closely replicating the properties of skin. Young's modulus of skin when measured by suction tests is reported to lie between 0.05 and 0.15 MPa, and indentation tests lie between 0.01 and 0.02 MPa \cite{pailler2008}. The modulus of our silicone block is catalogued at 0.06 MPa, making it a good candidate to replicate the effect of skin. We place the silicone block with a diameter of 66 mm and height of 32.5 mm, underneath each sample so the spirals rest on the soft block, as shown in Fig. \ref{fig:4}(a). We then perform a frequency sweep as described above and identify the first mode of resonance in each case. We notice that the resonance frequency of the samples with structure 1 (TMA) and structure 2 (no mass) on contact with the soft block is 69.01 Hz. This is very close to the resonance frequencies of the structures designed without the silicone contact. 

We repeat the experiment where we excite each structure in Fig. \ref{fig:4}(a) at their resonance frequencies for 15 seconds and note the displacements versus time at the center of each spiral. The Fourier transform of that data is shown in Fig. \ref{fig:4}(b) and (c) (blue dashed curve represents the TMA and red dashed curve represents the structure without a mass, in contact with the soft block). We compare their responses with the response of the same TMA vibrating with a free end (blue solid curve in Fig. \ref{fig:4}(b)). The TMA vibrates with a peak amplitude of 1.153 mm with a free end, and this drops to 0.231 mm when in contact with the soft block. This is well above the threshold of 0.020 mm that can be perceived by fingertips with a ball indenter at frequencies between 10 Hz and 100 Hz \cite{israr2007}. However, the structure without a mass vibrates at just 0.030 mm when in contact with the soft block, which is very close to the perceivable limit. This explains why we clearly feel the tuned mass vibrate with our fingertip whereas we hardly feel a vibration with the other structure.

In order to quantify the force and pressure feedback from each structure, we first measure the stiffness of soft block on an Instron machine as shown in Fig. \ref{fig:4}(d). Here, we utilize a larger indenter that has a circular cross-section with a diameter of 6.75 mm. We choose this to approximate the contact area of a fingertip with each structure as well as the contact area of each structure with the soft block at resonance (estimated from the mode shapes in Fig. \ref{fig:2}(c)) . The force versus displacement response from the small deformation quasi-static compression test at 0.01 mm/s is shown in Fig. \ref{fig:4}(e). Its small deformation stiffness of 395.20 N/m is measured as the slope of the curve near the origin. We can approximate the force from each structure in contact with the soft block as the product of peak displacements in Fig. \ref{fig:4}(c) with the measured stiffness of the block. This results in a peak force of 0.0915 N from the TMA (structure 1) and a force of 0.0115 N from the structure without a local mass (structure 2). This translates to a pressure of 2.558 KPa and 0.321 KPa in the structures with and without the mass, respectively.

\section{Discussion}
The TMA clearly displays a much higher stiffness when compared to the structure without a tuned mass, that resonates at the same frequency, as demonstrated by our analysis and experiments. This can be attributed to the spiral having a higher thickness (and stiffness) as well as the local stiffness contribution from both the spiral and cantilever in the TMA. The structure without the mass amplifies displacements purely by maximizing localization and compliance. Embedding a mass in the structure does not alter the stiffness as the mass is localized, but drastically drops the resonance frequency, hence allowing us to design high stiffness structures that resonate at low frequency. Further, we notice in experiments that the TMA resonates with a much higher displacement amplitude than its no mass counterpart. This can be attributed to the coupled resonance we design where the cantilever and spiral resonate in-phase at these low frequencies in the TMA.  

The properties of the structures that are designed and tested in this paper lie within the bounds estimated in Fig. \ref{fig:2}(b). Similarly, one could design structures with or close to other frequencies and stiffness within these bounds by appropriately modifying the objective in problem \eqref{eq:opt_samples}. Note that all attainable properties within the bounds on the design parameters in table \ref{tab:2} lie within the blue shaded region in Fig. \ref{fig:2}(b), but not all values in the region may be attained. This is because the bounds represent a convex hull of the possible values. However, all values where the tangents meet the shaded region in Fig. \ref{fig:2}(b) are attainable, and such extremal properties are often of interest.

We notice that the resonance frequency of the structures slightly increase when the spiral end of the cantilever is in contact with the soft block. This can be attributed to the stiffness increase due to the cantilever and soft block assembly, compared to just the cantilever in air. This increase would need to be taken into account when designing such amplification mechanisms against skin. Further, on contact with the soft block, we notice that the displacements drop when compared to a case without contact. We observe this due to the resistance to deformation provided by the silicone block. However, the amplitude at resonance of the TMA in contact with the block is 7.7 times the value when compared to structure 2, and this can be attributed to the higher stiffness of the TMA. Finally, the forces and pressures from the structure can be amplified by increasing the local stiffness of the TMA. This can be done in a couple of ways. We can increase the thickness of the spiral and place a heavier mass, or we can print the TMA with a stiffer base material and use a heavier mass to maintain the same resonance frequency.

\section{Conclusion}

We present a tuned mass amplifier (TMA), that is capable of enhancing haptic feedback by amplifying local displacements together with local stiffness and low frequencies suitable for haptics. We amplify the local displacements by utilizing coupled resonance, where a primary structure (cantilever) resonates in-phase with a secondary structure (compliant spiral cantilever patch). The embedded mass helps drop the frequency of resonance to low values, while still maintaining high structural rigidity. We establish an optimal design framework by first learning the finite element simulations given the design parameters using a deep neural network. We then calculate the bounds on attainable mechanical properties using a sensitivity analysis. Within these bounds on stiffness and resonance frequency, we design the parameters describing the geometry to achieve prescribed mechanical properties. We compare two structures- structure 1 is a TMA that features a tuned mass and resonates at 65 Hz, and structure 2 does not have an embedded mass and resonates at the same frequency. We validate the high stiffness and displacement amplification seen in the TMA through experiments on 3-D printed structures made of Nylon-12 with an embedded Tungsten Carbide mass. We also conduct experiments with the structures in contact with a soft silicone block to quantify the effect of touch in haptics. The structures can be made of arbitrary geometries such as a curved cantilever for remote haptics, where the site of vibration actuation is further away from the location of haptic feedback. Also, the spiral patch can be engineered as a network of truss-like connections or a soft continuous material embedding a mass to explore higher amplifications. The proposed designs can be made arbitrarily small to realize such phenomena in wearable haptics.

\section*{Data availability}

The data that support the findings of this study can be made available from the corresponding author upon reasonable request.

\section*{Acknowledgements}

We thank Hamzeh Musleh Fahmawi and Joseph Aase for help with conduting the experiments. We also thank Dr. Maurizio Chiaramonte and Dr. Kevin Carlberg for useful discussions.

\bibliographystyle{unsrtnat}
\bibliography{main}
\end{document}